\documentstyle[11pt,paspconf,epsf]{article}

\newcommand {\hI} {H I\,\,}
\newcommand {\hII} {H II\,\,}
\newcommand {\ha} {H$\alpha$\,\,}
\newcommand {\kms} {\,km\,s$^{-1}$\,}
\newcommand {\M} {\mbox{${\cal M}$}}

\newcommand {\msol} {\M$_\odot$\,}

\newcommand {\mlb} {(\M/L$_B$)$_\star$\,\,}

\newcommand {\mlsol}{\mbox{${\cal M}_\odot$/L$_{\odot}$}}
\newcommand {\mhIhe} {\M$_{HI+He}$\,}
\newcommand {\mldyn} {(\M/L$_B$)$_{dyn}$\,\,}

\begin{document}

\title{Necessity for High Accuracy Rotation Curves in Spiral Galaxies}

\author{S\'ebastien Blais--Ouellette}
\affil{D\'epartement de physique and Observatoire du mont M\'egantic,
Universit\'e de Montr\'eal, C.P. 6128, Succ. centre ville,
Montr\'eal, Qu\'ebec, Canada. H3C 3J7
and IGRAP, Observatoire de Marseille,
2 Place Le Verrier, F--13248  Marseille Cedex 04, France\\
e--mail: blaisous@astro.umontreal.ca}

\author{Claude Carignan} 
\affil{D\'epartement de physique and Observatoire du mont M\'egantic,
Universit\'e de Montr\'eal.\\
e--mail: carignan@astro.umontreal.ca}

\author{Philippe Amram}
\affil{IGRAP, Observatoire de Marseille.\\
e--mail: amram@observatoire.cnrs-mrs.fr}

\begin{abstract}
In the last 20 years, rotation curves derived from \hI kinematics
obtained on radio synthesis instruments were used to probe the dark
matter distribution in spiral and dwarf irregular galaxies. It is shown,
with the aid of the Sd galaxy NGC 5585, that high 
resolution 2--D \hII kinematics is necessary to determine accurately
the mass distribution of spirals. 
New CFHT Fabry--Perot \ha observations are combined with low resolution  
Westerbork \hI data to study its mass distribution. 
Using the combined rotation curve and best fit models,
it can be seen that the \mlb of the luminous disk goes
from 0.3, using only the \hI rotation curve, to 0.8, using both the
optical and the radio data. This reduces the dark--to--luminous mass ratio 
by $\sim 30$\%.
\end{abstract}

\keywords{cosmology: dark matter --- galaxies: individual (NGC 5585)\\
--- galaxies: fundamental parameters (masses) --- techniques: interferometry}

\section{Introduction}

Is has often been argued that since \hI observations probe the
gravitational potential well outside the optical radius, in the dark matter 
dominated region, they are best suited to derived the characteristics of the 
mass distribution, thus of the {\bf dark} mass distribution. However, this is 
skipping the fact that the parameters of the mass models (and especially of 
the DM distribution) are very sensitive, as shown below, to the
rising inner part of the rotation curve which can be derived with greater 
precision using 2--D \ha observations (see e.g. Amram et al. 1996).

\begin{figure}
\plottwo{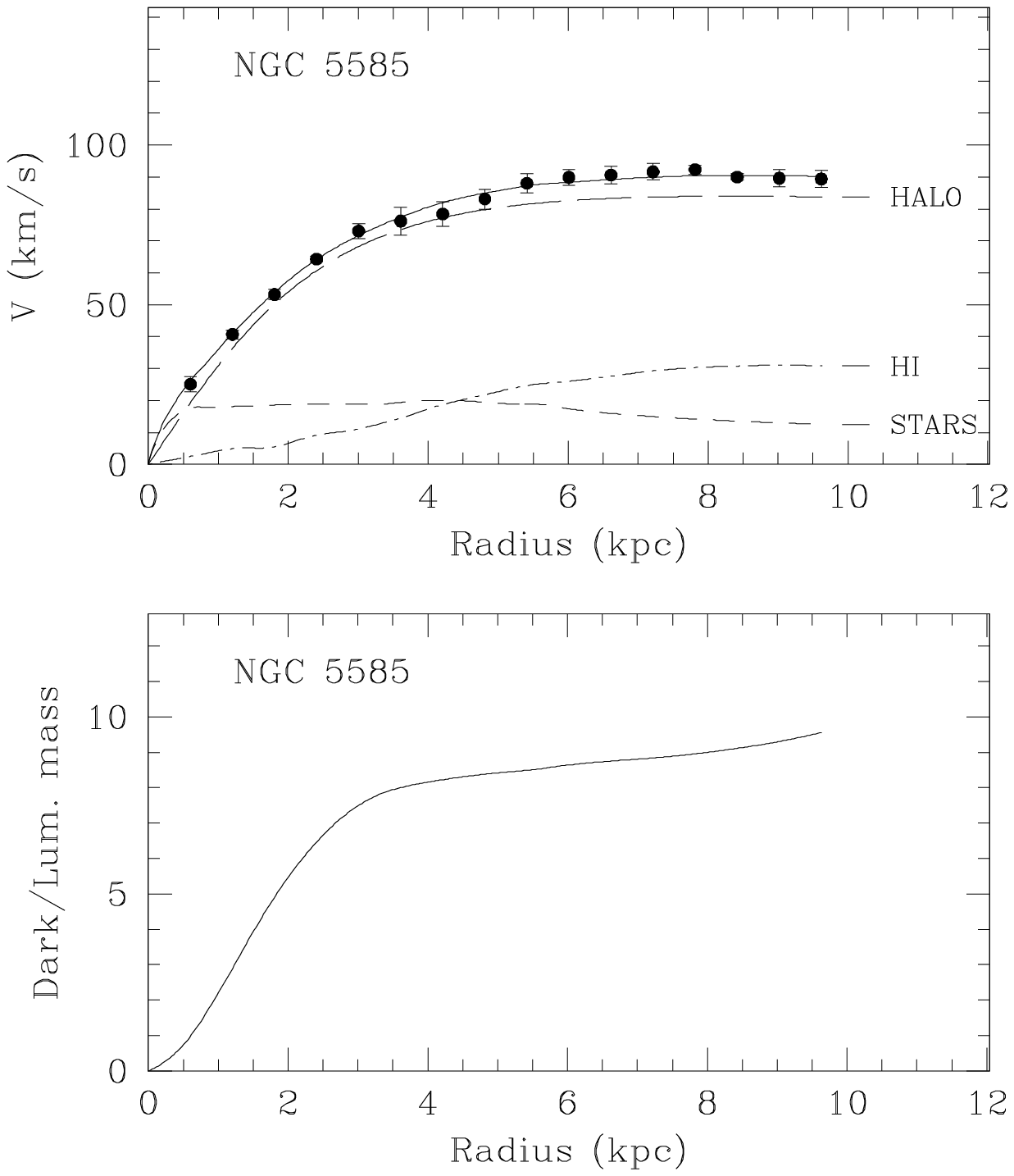}{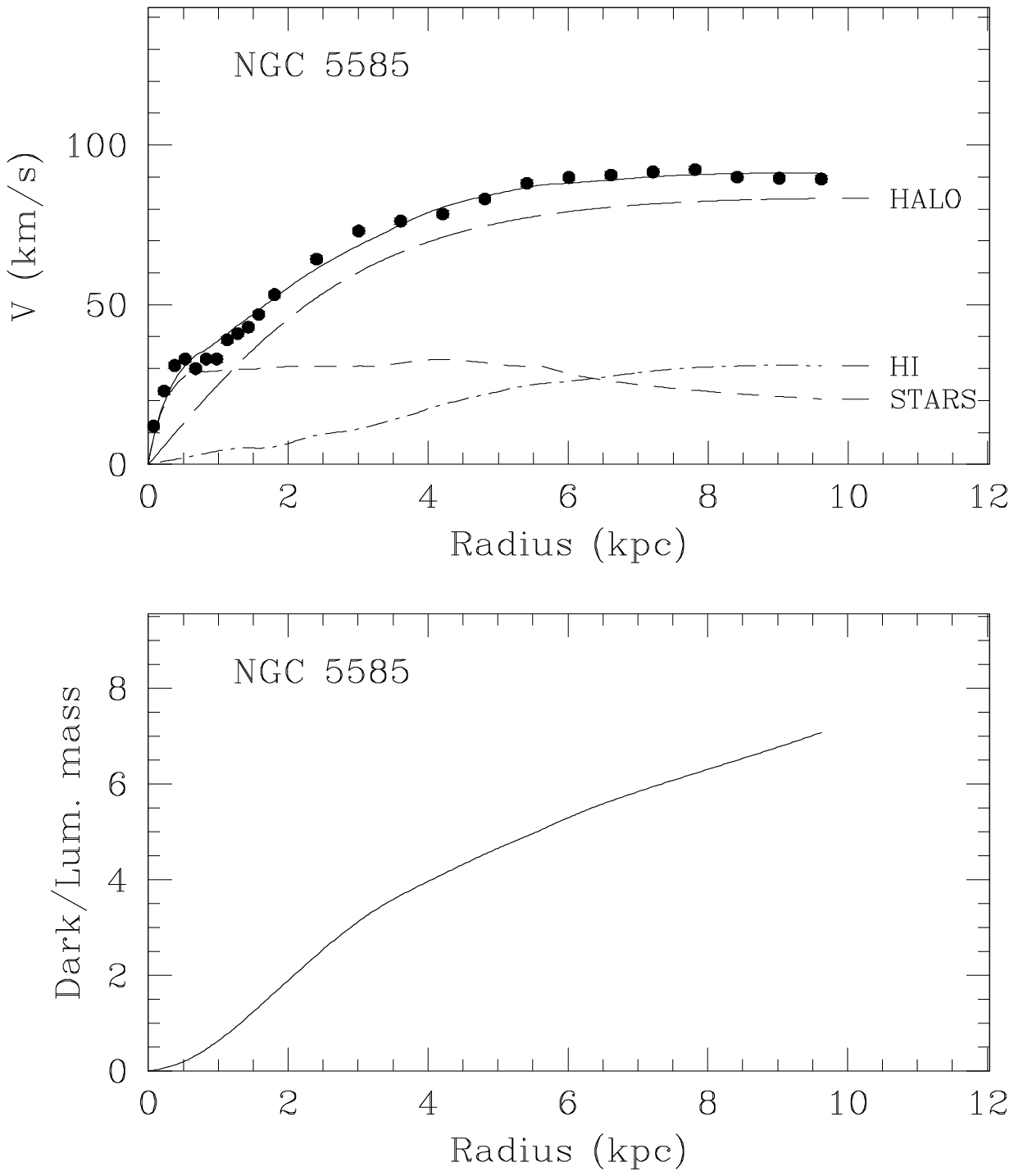}
\caption{{\bf a)} Best fit mass model for NGC 5585 using the \hI rotation curve 
(left) and \ha \& \hI (right).
{\bf b)} Dark--to--luminous mass ratio as a function of radius.\label{fig1}}
\end{figure}

The left part of figure \ref{fig1}, which shows the mass distribution of NGC 5585 
using its \hI 
rotation curve, illustrates the now commonly used method to study the mass 
distribution (Carignan \& Freeman 1985).
First, the rotation curve is obtained by fitting  a 
``tilted--ring'' model to the \hI velocity field in order to represent the
warp of the \hI disk which is very often present. The accuracy of the
model representation is then checked by looking at the residuals
(data - model) map. Then, the
luminosity profile, preferentially in the red, is transformed into
a mass distribution for the stellar disk assuming a constant \mlb
(Casertano 1983, Carignan 1985). For the contribution of the gaseous
component, the \hI radial profile is used, scaled by 4/3 to account
for He. The difference between the observed rotation curve and the
computed contribution of the luminous (stars \& gas) component is thus
the contribution of the dark component which can be represented by
an isothermal halo (Carignan 1985) or some other functional form.

The \hI observations, often optimized for maximum
sensitivity in the outer parts, have in most of the published studies
a resolution of only 20--45 \arcsec. The example of NGC 5585 will show that 
this is clearly not sufficient to well determine the rising part of the 
rotation curve and that this part is crucial since it strongly constrains the 
three free parameters of the mass model (see also Swaters et al. in these 
proceedings).

\section{Observations and data analysis}

The Fabry--Perot observations of the \ha emission line were obtained
in 1994 February at the Canada--France--Hawaii Telescope (CFHT).The Fabry--Perot
etalon (CFHT1) was installed in the CFHT's Multi--Object Spectrograph (MOS).
A narrow--band filter ($\Delta \lambda$ = 10\,\AA), centered at
$\lambda_0$ = 6570\,\AA\, (nearly at the systemic velocity of NGC
5585, V$_{sys} \approx 305$ \kms), was placed in front of the etalon.
The available field with no vignetting was $\approx$ 8.7\arcmin $\times$
8.7\arcmin, with .34\arcsec\, pix $^{-1}$. The free spectral range of
5.66\,\AA\, (258 \kms) was scanned in 28 channels, giving a sampling
of 0.2\,\AA\, (9.2 \kms)per channel. Eight minutes integration was
spent at each channel position. 

After standard reduction, images are assembled in a cube and phase calibrated in 
order to get one particular wavelength in a plane. A velocity map is then 
obtained using the intensity weighted mean of the H$\alpha$ peak in each pixel.

The rotation curve have been deduced from the velocity field following
two different methods. The first estimate was made using ROCUR (Begeman
1987) where annuli in the plane of the galaxy (ellipse in the plane
of sky) are fitted to the velocity field in order to minimize the 
velocity dispersion inside these rings. This allows to determine the
rotation center, systemic velocity, position angle and inclination.

The ADHOC package (Boulesteix 1993) have then been used to fine tuned these 
parameters by
direct visualization and comparison with residual velocity field. 
Note that the internal errors on \hII rotation
curves data points are usually larger than on \hI rotation curves
data points reflecting the larger local motions in the \hII regions.

\section{Mass models}

For our adopted mass model of NGC 5585, we combine the high
resolution \ha data in the inner parts high sensitivity
of the \hI data in the outer parts. Since we are doing best--fit models,
we have to be careful. First, as explained earlier, the \hII data points
have larger intrinsic errors.
This means that they will carry smaller weights in the fitting process.
On the other hand, because of the higher resolution, there is more \hII
data points than \hI data points. This means that the optical data
would tend to have a higher weight than the radio data. To avoid any
bias, we decided to use for the final model the \ha data for r~$<$~120\arcsec\,
and the \hI data for r~$>$~120\arcsec. Table~\ref{tab:mod5585}
gives the parameters of the mass models constructed using
only the \hI rotation curve, the \ha rotation curve,
and those for the model using the combined
\hI \& \ha rotation curve.
 
This adopted model is showed in the right part of figure \ref{fig1}. 
As expected, $\sigma$ is very similar in the combined \hI \& \ha
curve than in the \hI rotation curve. This is the case because
this parameter is a measure of the maximum amplitude of
the rotation curve which is mainly defined by the \hI data in the
outer parts. However, the two other parameters \mlb for the stellar
disk and r$_c$ of the dark halo (which are coupled) have nearly the
same values than those derived with the \ha curve. Again, this is because the
\mlb of the luminous stellar disk, and hence the scaling parameter
of the dark halo r$_c$, are mainly constrained by the \hII data in
the inner parts.  
 
In summary, adding the \ha data to the \hI data of NGC 5585, increases
in the best--fit model the \mlb of the stellar disk from 0.3 to 0.8.
The result is to push the dark halo outward and reduced by
$\sim$ 30\% the dark--to--luminous mass ratio.
 
\section{Conclusions}
 
The importance of an accurate determination of the rising part of
the rotation curves using full 2--D high resolution Fabry-Perot
observations is well illustrated by the example of NGC 5585 which
resulted in a $\sim$30\% difference in its dark--to--luminous mass
ratio. This is one of the motivation behind the {\bf GHASP} project, 
which intends to map accurately
the \ha velocity fields of $\sim$200--300 nearby spiral and dwarf
galaxies using Fabry--Perot observations. The candidate
galaxies will be selected from the {\bf WHISP} project, which intend to
map the \hI in 
1000--3000 galaxies in the next ten years. This should be the best way
to study possible correlations between the parameters of the DM
halos and other properties of the galaxies.
 
\begin{table}
\caption{Parameters of the mass models of NGC 5585.\label{tab:mod5585}}
\begin{center}\scriptsize
\begin{tabular}{l|ccc}
\tableline
{\bf Parameter}&{\bf \hI RC}&{\bf \ha RC}&{\bf Combined \hI \& \ha RC}\\
{\it Luminous disk component}:                                            \\
\mlb    \hfill  (\mlsol) &0.3 $\pm 0.3$ \tablenotemark{a} &0.9 $\pm 1.5$ &0.8 $\pm 0.2$  \\
\M$_\star$ \hfill (\msol)       &$3.3 \times 10^8$      &$9.9 \times 10^8$ &$8.8 \times 10^8$\\
\mhIhe\hfill    (\msol)         &$1.4 \times 10^9$      &$1.4 \times 10^9$ &$1.4 \times 10^9$\\
\\
{\it Dark halo component}:                                                \\
r$_c$ \hfill    (kpc)           &2.8 $\pm 0.3$          &3.8 $\pm 2.5$ &3.9 $\pm 0.4$     \\
$\sigma$ \hfill (\kms)          &52.9 $\pm 2.0$         &48.0 $\pm 15.0$ &53.3 $\pm 1.5$          \\
$\rho_0$ \hfill (\msol pc$^{-3}$)&0.060         &0.027 &0.031             \\
\\
{\it At R$_{HO}$ r = 6.5 kpc}:                                            \\
$\rho_{halo}$ \hfill (\msol pc$^{-3}$)&0.0035           & &0.0041                 \\
\M$_{dark+lum}$ \hfill (\msol) &$1.2 \times 10^{10}$ & &$1.1 \times 10^{10}$\\
\mldyn                          &10.6                   & &10.3           \\
\M$_{dark}$/\M${lum}$           &8.7                    & &5.5            \\
\\
{\it At the last measured point r = 9.6 kpc}:                             \\
$\rho_{halo}$ \hfill (\msol pc$^{-3}$)&0.0013           & &0.0016                 \\
\M$_{dark+lum}$ \hfill (\msol) &$1.7 \times 10^{10}$ & &$1.8 \times 10^{10}$\\
\mldyn                          &10.6                   & &10.3           \\
\M$_{dark}$/\M${lum}$         &9.5                      & &7.2            \\
\end{tabular}
\end{center}
\end{table}

\end{document}